\documentstyle[aps,11pt,psfig,preprint]{revtex}
\newcommand{\peta}{p_{\eta}}
\begin{document}
\draft
\title{Dependence of calculated binding energies and widths
of $\eta$-mesic\\ 
nuclei on treatment of subthreshold $\eta$-nucleon interaction }
\author{ Q. Haider }
\address{Physics Department, Fordham University,
Bronx, N. Y. 10458}
\author{ L.C. Liu }
\address{Theoretical Division, Los Alamos National Laboratory,
Los Alamos, N. M 87545}
\maketitle

\begin{abstract}
We demonstrate that the binding energies $\epsilon_{\eta}$ and 
widths $\Gamma_{\eta}$ of $\eta$-mesic nuclei 
depend strongly on subthreshold $\eta$-nucleon interaction.
This strong dependence is made evident from comparing 
three different $\eta$-nucleus 
optical potentials:   
(1) a microscopic optical potential taking into account the full effects   
of off-shell $\eta N$ interaction; (2) a factorization approximation
to the microscopic optical potential where
a downward energy shift parameter is introduced      
to approximate the subthreshold $\eta N$  
interaction; and (3) an optical potential using         
on-shell $\eta N$ scattering length  as the interaction input. 
Our analysis indicates that the in-medium $\eta N$ interaction for
bound-state formation is about 30~MeV below the free-space $\eta$N
threshold, which causes a substantial reduction of the 
attractive force between the $\eta$ and nucleon with respect to that  
implied by the scattering length. Consequently, 
the scattering-length approach overpredicts the $\epsilon_{\eta}$ 
and caution must be exercised when these latter predictions are used       
as guide in searching for $\eta$-nucleus bound states.   
We also show that final-state-interaction analysis cannot provide 
an unequivocal determination of the existence of $\eta$-nucleus bound 
state. More direct measurements are, therefore, necessary.  
\end{abstract}


\narrowtext
\section{Introduction}\label{sec:1}
The existence of $\eta$-mesic nucleus, a bound state of $\eta$ meson
in a nucleus, was first predicted by us in 1986~\cite{hai1}. The
formation of such bound systems is caused by the attractive interaction
between the $\eta$ meson and all the nucleons in the nucleus. The
attractive nature of the interaction followed naturally from the
work of Bhalerao and Liu~\cite{bhal} who found, from a detailed
coupled-channel analysis of $\pi N\rightarrow \pi N$,
$\pi N\rightarrow \pi\pi N$, and $\pi N\rightarrow \eta N$ reactions,
that near-threshold $\eta N$ interaction is attractive. We can easily 
understand this attraction by observing that the $\eta N$ threshold is 
situated just below the $N^{*}(1535)$ resonance.  However,
the attraction given in ref.\cite{bhal} is not strong enough 
to support a bound state
in nuclei with mass number $A<12$. This latter conclusion\cite{hai1}
was confirmed by Li et al.~\cite{liku}, who used a standard
Green's function technique of many-body theory to study the formation
of $\eta$-mesic nuclei.
Before proceeding further, we would like to mention that in the
literature the $\eta$-nucleus bound states are also called $\eta$-nucleus
quasibound states. This is because these bound states have
a finite width and they eventually decay.
In this work,  we shall simply call them bound states.

If the existence of $\eta$-mesic nucleus
is experimentally confirmed, many new studies of  
nuclear and particle physics will become possible.
For example, because the binding energies of $\eta$ meson
depend strongly on the coupling between the $\eta N$ and the $N^{*}(1535)$
channels~\cite{bhal}, studies of $\eta$-mesic nucleus can yield additional 
information on the $\eta NN^{*}$ coupling constant involving bound 
nucleons. Furthermore, the
$\eta$-mesic nuclear levels correspond to an excitation energy of
$\sim 540$ MeV, to be compared with an excitation energy of
$\sim 200$ MeV associated with the $\Lambda$- and $\Sigma$-hypernuclei.
The existence of nuclear bound states with such high excitation
energies provides the possibility of 
studying nuclear structure far from equilibrium. 

Several experiments~\cite{chri,lieb,john} using $\pi^{+}$
beam, as motivated by the theoretical works of refs.~\cite{hai2,hai3}, 
were performed to search for $\eta$-mesic nuclei. While these
experiments could not confirm the existence of $\eta$-mesic
nucleus, they did not rule out such possibility either. More recently,  
Sokol et al.~\cite{sok1,sok2} have claimed observing the
$\eta$-mesic nucleus $^{11}_{\eta}$C through measuring the invariant mass 
of correlated $\pi^{+}n$ pairs in a photo-mesonic reaction. Further 
confirmation of their experimental result, with improved statistics,
is needed before arriving at a definite conclusion.

The existence of these $\eta$-mesic nuclear states depends on the value of  
the $\eta N$ scattering length $a_{\eta N}$. (For the scattering length,
we use the sign convention of Goldberger and
Watson~\cite{wats}.)  After 1990, 
many theoretical models were proposed for $\eta N$ interaction. Fits to  
various data have yielded very different $a_{\eta N}$ 
\cite{bhal,gre3,gre2,saue,benn,abae,kais,bat1,bat3,bat2,arim,tiat,krus}.
The real part of $a_{\eta N}$ ranges from
as low as 0.270~fm~\cite{bhal} to as high as 1.000~fm~\cite{gre2},
while the imaginary part varies between 0.190~fm~\cite{bhal}
to 0.399~fm~\cite{krus}. This  
wide range of values of $a_{\eta N}$
are summarized in table~\ref{table:1}.
These very different values arise because  
$a_{\eta N}$ is not directly measurable and
must be inferred indirectly from other observables, such as the $\pi N$ 
phase shifts. In this latter case, the calculated $a_{\eta N}$ depends
strongly on the model used to relate the $\eta N$ to the $\pi N$ channels.   
As pointed out in ref.~\cite{bat3},  
the inclusion of both the $N^{*}(1535)$ and $N^{*}(1650)$ 
resonances leads to larger scattering lengths. However, this general rule
does not account for all the reasons.  
For example, in ref.\cite{abae} the inclusion of both
these $N^{*}$ resonances, but with fit to different reaction data, does not
lead to a ${\cal R}e~a_{\eta N}$ as high as  $\approx 0.8$ fm rather only
to a ${\cal R}e~a_{\eta N}\approx 0.6$ fm.
We caution the readers that it
is premature to conclude which published $a_{\eta N}$ is the realistic one.
Such a determination can be made only after $\eta$-mesic nuclei are
experimentally discovered, as only their experimental observation can
set a stringent constraint on the in-medium scattering length and enable
one to differentiate between different theoretical models.

Using final-state-interaction (FSI) analysis, several
theoretical studies have suggested an $a_{\eta N}\approx (0.5 + 0.3i)$~fm
and the possibility of the formation of
bound $\eta$-nucleus states in light nuclei, 
such as $^{3,4}$He~\cite{wilk,wyce} and deuteron~\cite{gre1,rak1,rak2}. 
We will revisit the FSI analysis in this work and show the limitation 
of the method.   In spite of the  
fact that many predicted $a_{\eta N}$ (see table~\ref{table:1}) are much  
larger than the one used in the above-mentioned FSI analysis,   
to this date there has been no direct observation of these bound states 
in light nuclei. Consequently, we believe it is valuable  
to understand the current experimental situation by analyzing in detail  
the dynamics pertinent to the formation of $\eta$-mesic nucleus. 

We have learned from studying pion-nucleus scattering that,  
because of the Fermi motion and binding of the target nucleon, 
the pion-nucleon interaction in nuclei occurs at energies below its  
free-space value, i.e., below the energy that the pion-nucleon system   
would have when the nucleon is unbound\cite{liu1}.
As the above causes for the lowering of the interaction energy 
are also present in $\eta$-nucleus scattering, we believe that the  
$\eta$-nucleon interaction energy in $\eta$-mesic nucleus formation will occur 
below its free-space threshold and this subthreshold interaction 
can have important effects on  
the formation of $\eta$-nucleus bound states.  
This observation has motivated us to examine in this work  
the reliability of using the free-space $\eta N$ information, such as the 
$\eta N$ scattering length, to make predictions of  
the nuclear binding of the $\eta$.   
 
We will, therefore,  study the dependence of $\eta$-nucleus bound-state
formation on off-shell behavior of $\eta N$ interaction in nuclei 
with $A\geq 3$ through comparing three different  
theoretical approaches. They are: (1) a fully off-shell microscopic
optical potential that uses an off-shell $\eta N$ model;  
(2)~a factorization approximation (FA) to the preceeding microscopic
potential, where an energy shift parameter is used to calculate the
$\eta N$ interaction at subthreshold energies; and (3)
an ``on-shell'' optical potential whose strength depends solely on
the scattering length $a_{\eta N}$. The last approach has its root     
in the study of K- and $\pi$-mesic atoms~\cite{koch,seki,kwon} and is 
also extensively used in predicting $\eta$-mesic nuclei.
Finally, we mention that as the existence of $\eta$-mesic nucleus has not
yet been experimentally confirmed with certainty, there is no data
to be fitted. Hence, all of our results as well as those of others are
purely predictions.

The paper is organized as follows. The three theoretical approaches are
described in sec.~\ref{sec:2}. The important differences in their underlying 
reaction dynamics are outlined. Results and discussion are
presented in sec.~\ref{sec:3}. 
As we shall see, for a given $\eta N$ interaction model, 
the ``on-shell'' approach gives much stronger binding 
for $\eta$ than does the microscopic calculation, making evident  
the important effects of off-shell dynamics. We will further show that   
these effects are of very general nature and are independent of the specific
unitary off-shell model used in the comparative study.  
The ramifications of off-shell effects on both experimental and theoretical
studies of bound states of $\eta A$ systems, particularly in systems
with $3\leq A <12$, are discussed. In view of the interest in the method 
of FSI, we also reexamine in more detail
this method and its applications to the study of
$\eta$-$^{3,4}$He systems. The conclusions of our study are summarized in 
sec.~\ref{sec:4}.

\section{Theory}\label{sec:2}
We calculate the binding energy $\epsilon_{\eta}$ and width
$\Gamma_{\eta}$ of an $\eta$-nucleus bound system by solving
the momentum-space relativistic three-dimensional
integral equation 
\begin{equation}
\frac{{\bf k'}^{2}}{2\mu}\;{\psi}({\bf k'})
+ \int\;d{\bf k}<{\bf k}'\mid {V}\mid{\bf k}>{\psi}({\bf k})
=E{\psi}({\bf k}')\ , 
\label{eq:2.1}
\end{equation}
using the inverse-iteration method~\cite{kwon}. 
Here $<{\bf k'}\mid V \mid {\bf k}>$ are  
momentum-space matrix elements of the $\eta$-nucleus
optical potential $V$, with  
${\bf k}$ and ${\bf k}'$ denoting, respectively, 
the initial and final $\eta$-nucleus relative momenta.  
The $\mu$ is the reduced mass of the $\eta$-nucleus system and 
$E$ is the complex eigenenergy which we will 
denote as $ \epsilon_{\eta}+i\Gamma_{\eta}/2 \equiv 
\kappa^{2}/2\mu.$
For bound states, $\epsilon_{\eta} < 0$ and $\Gamma_{\eta} < 0.$
As mentioned in sec.~\ref{sec:1}, 
three different approaches to $V$ are used by us to calculate $E$, 
and they are described below. 

\subsection{Covariant $\eta$-nucleus optical potential}\label{sec:2.1}
In spite of its Schr\"{o}dinger-like form, eq.(\ref{eq:2.1}) is covariant
as it can be obtained from applying a specific covariant 
reduction\cite{cel1} to the relativistic bound-state 
equation $ \Psi = G_{0}{\cal V}\Psi $.  
The three-dimensional relativistic wave function $\psi$ and
the covariant potential $<{\bf k}'\mid V\mid~{\bf k}>$ in 
eq.(\ref{eq:2.1}) are related to the fully relativistic ones by   
\begin{equation}
\psi({\bf k}) = \sqrt{ \frac{R(\kappa_{r}^{2})}{R({\bf k}^{2})}}\; 
\Psi ({\bf k},k^{0})\ ,
\label{eq:2.1.1}
\end{equation}
and
\begin{equation}
<{\bf k}'\mid{V}\mid{\bf k}> = \sqrt{R({\bf k}'^{2})}
<{\bf k}'\mid {\cal V}(W,k'^{0},k^{0})\mid{\bf k}>\sqrt{R({\bf k}^{2})}\ .
\label{eq:2.1.2}
\end{equation}
In the above equations,
\begin{equation}
W=\sqrt{M_{\eta}^{2}+\kappa_{r}^{2}}+\sqrt{M_{A}^{2}+\kappa_{r}^{2}}\ ; 
\ \ \kappa_{r}^{2}\equiv 2\mu\epsilon_{\eta}\ , 
\label{eq:2.1.3}
\end{equation}
and 
\begin{equation}
R({\bf k}^{2}) = \frac{M_{\eta}+M_{A}}{E_{\eta}({\bf k})+
E_{A}({\bf k})}\ .
\label{eq:2.1.4}
\end{equation} 
However, as a result of the application of the covariant reduction, 
the zero-th components of the four-momenta ${\bf k}$ and ${\bf k}'$ are no 
longer independent variables but are constrained by 
\begin{equation}
k^{0}=W-E_{A}({\bf k}), \;\;\;
k'^{0}=W-E_{A}({\bf k}')\ .
\label{eq:2.1.5}
\end{equation}
The main advantage of working with a covariant theory is
that the $\eta$-nucleus interaction $V$ can be related to the
elementary $\eta N$ process by unambiguous kinematical
transformations~\cite{liu1}.

The first-order microscopic $\eta$-nucleus optical potential
has the form 
\begin{eqnarray} 
 <{\bf k}'\mid V\mid{\bf k}> & = & \sum_{j}\int d{\bf Q}
<{\bf k}',-({\bf k}'+{\bf Q})\mid t(\sqrt{s_{j}})_{\eta N\rightarrow\eta N}
\mid {\bf k}, -({\bf k}+{\bf Q})> \nonumber \\   
 & \times & \phi^{*}_{j}(-{\bf k}'-{\bf Q})
\phi_{j}(-{\bf k}-{\bf Q})\ ,
\label{eq:2.1.6}
\end{eqnarray}
where the off-shell $\eta N$ interaction  
$t_{\eta N\rightarrow\eta N}$ is weighted by the product of the nuclear
wave functions $\phi^{*}_{j}\phi_{j}$ corresponding to having the 
nucleon $j$ at the momenta $-({\bf k}+{\bf Q})$ and $-({\bf k}'+{\bf Q})$
before and after the collision, respectively. The $\sqrt{s_{j}}$ is
the $\eta N$ invariant mass and is equal to the total energy in the
c.m. frame of the $\eta$ and the nucleon $j$. It is given by\cite{liu1} 
\begin{eqnarray}
s_{j} & = & [\{ W-E_{C,j}({\bf Q})\}^{2}-{\bf Q}^{2}] \nonumber \\  
 & \simeq & 
\left [ M_{\eta}+M_{N}-\mid\epsilon_{j}\mid \; -\; \frac{{\bf Q}^{2}}
{2M_{C,j}}\;\left ( \frac{M_{\eta}+M_{A}}{M_{\eta}+M_{N}} \right )
\right ]^{2} < (M_{\eta}+M_{N})^{2}\ ,
\label{eq:2.1.7}
\end{eqnarray}
where ${\bf Q}$,  $E_{C,j}$ and $M_{C,j}$ are, respectively, the 
momentum, total energy, and mass
of the core nucleus arising from removing a nucleon $j$ of 
momentum $-({\bf k}+{\bf Q})$ and binding energy $\mid\epsilon_{j}\mid$
from the target nucleus having the momentum $-{\bf k}$. 
Equations (\ref{eq:2.1.6}) and (\ref{eq:2.1.7}) indicate that the
calculation of $V$ involves integration over the
Fermi motion variable ${\bf Q}$ and requires knowledge of the basic
$t_{\eta N\rightarrow\eta N}$ at
subthreshold energies. 

The matrix element of $t_{\eta N\rightarrow\eta N}$ in the $\eta$-nucleus 
system is related to
the $\eta N$ scattering amplitude $\cal A$ in the $\eta N$ system by
\begin{eqnarray} 
 <{\bf k}',-({\bf k}'+{\bf Q}) & \mid & 
t(\sqrt{s_{j}})_{\eta N\rightarrow \eta N}\mid
{\bf k},-({\bf k}+{\bf Q})> \nonumber \\   
& = & 
\frac{\sqrt{E_{\eta}({\bf p}')
E_{N}({\bf p}')E_{\eta}({\bf p})E_{N}({\bf p})}}
{\sqrt{E_{\eta}({\bf k}')
E_{N}({\bf k'+Q})E_{\eta}({\bf k})E_{N}({\bf k+Q})}} {\cal A}(\sqrt{s_{j}},
{\bf p'},{\bf p}) \ , 
\label{eq:2.1.9}
\end{eqnarray}
where ${\bf p}$ and ${\bf p}'$ are the initial and final relative
three-momenta in the $\eta N$ c.m. frame.
We define the on-shell limit as 
$p'= p =p_{o}$ and $\sqrt{s_{j}} = E_{\eta}(p_{o}) + 
E_{N}(p_{o})\equiv \sqrt{s_{o}}$,
where $p_{o}$ is the on-shell (asymptotic) momentum. 
A natural way of parameterizing ${\cal A}$ is
\begin{equation}  
 {\cal A}(\sqrt{s_{j}},{\bf p'},{\bf p}) =
-\;\frac{\sqrt{s_{j}}}{4\pi^{2}\sqrt{E_{\eta}(p')  
E_{N}(p')E_{\eta}(p)E_{N}(p)} }  
 \;{\cal F}(\sqrt{s_{j}},{\bf p'},{\bf p})\ , 
\label{eq:2.1.10}
\end{equation}
so that $(d\sigma/d\Omega)_{\eta N\rightarrow \eta N} =
\mid {\cal F}\mid^{2}$. The ${\cal F}$ has the standard partial-wave 
expansion of a spin 0-spin 1/2 system: 
\begin{eqnarray}
{\cal F}(\sqrt{s_{j}},{\bf p'},{\bf p}) &=&   
 \frac{1}{\sqrt{p'p}}
   \sum_{\ell} [\  \left( \ell\ t^{\ell}_{2T,2j_{-}}(\sqrt{s_{j}},p',p)  + 
(\ell+1)t^{\ell}_{2T,2j_{+}}(\sqrt{s_{j}},p',p) \ \right)\;P_{\ell}(z)
\nonumber \\
 & - & i \vec{\sigma}\cdot (\hat{\bf p}\times \hat{\bf p'}) 
\left( t^{\ell}_{2T,2j_{-}}(\sqrt{s_{j}},p',p) 
   - t^{\ell}_{2T,2j_{+}}(\sqrt{s_{j}},p',p)\ \right)\;P_{\ell}'(z)\ ]\ ,        
\label{eq:2.1.10a}
\end{eqnarray}
where $z=\hat{\bf p}\cdot\hat{\bf p'}$,   
$j_{\pm}=\ell\pm 1/2$, and  $T$ is the  
isospin of the $\eta N$ system and equals to 1/2.    
In the on-shell limit,
\begin{equation} 
\frac{t^{\ell}_{2T,2j_{\pm}}(\sqrt{s_{j}},p,p)}{\sqrt{p'p}} \ 
{\longrightarrow}\ \frac{1}{2ip}  
 \left(\exp[\ 2i\delta^{\ell}_{2T,2j_{\pm}}(\sqrt{s_{o}})\ ] 
-1\right) \ . 
\label{eq:2.1.10c} 
\end{equation} 
The phase shifts $\delta^{\ell}$ are complex-valued  
because the thresholds for $\eta N\rightarrow\pi N$ and
$\eta N\rightarrow\pi\pi N$ reactions are lower than the threshold
for $\eta N$ scattering.   
When $p\rightarrow 0$, $\delta^{\ell}\rightarrow p^{2\ell+1}a^{(\ell)}$ and  
\begin{equation}  
\frac{t^{\ell}_{2T,2j_{\pm}}(\sqrt{s_{j}},p,p)}{\sqrt{p'p}} \ 
{\longrightarrow}\ p^{2\ell}a^{(\ell)}_{2T,2j_{\pm}}\ . 
\label{eq.2.1.10d} 
\end{equation}   
The $a^{(0)}_{2T,2j}$ and $a^{(1)}_{2T,2j}$ are, respectively,
the (complex) $\eta N$ scattering length and volume.  
Near the threshold,  only the $s$-wave term,  
$t^{0}_{11}$ in eq.(\ref{eq:2.1.10a}), is important.  

Different off-shell models give different off-shell extensions of $\cal A$ 
to kinematic regions
where $p \neq p'$ and $\sqrt{s_{j}}\neq \sqrt{s_{o}}$ .
In the separable model of ref.\cite{bhal}, the off-shell amplitude is
given by
\begin{equation}
t_{\alpha}(\sqrt{s_{j}}, p', p) = K(\sqrt{s_{j}},p',p)\ \sqrt{p'p}\  
\left ( \frac{N_{\alpha}(\sqrt{s_{j}},p',p)}{D_{\alpha}
(\sqrt{s_{j}})} \right )\ ,
\label{eq:2.1.12}
\end{equation} 
with
\begin{equation}  
 K =\ -\frac{\pi}{\sqrt{s_{j}}} 
\sqrt{E_{\eta}(p')E_{N}(p')E_{\eta}(p)E_{N}(p)}  \ ,
\label{eq:2.1.12a} 
\end{equation} 
\begin{equation} 
N_{\alpha}= h_{\alpha}(\sqrt{s_{j}},p')h_{\alpha}(\sqrt{s_{j}},p) 
\propto 
\frac{g^{2}_{\eta N\alpha}}{2\sqrt{s_{j}}}\ (p'p)^{\ell}v_{\ell}(p')v_{\ell}(p) \ ,  
\label{eq:2.1.12b} 
\end{equation} 
and 
\begin{equation} 
D_{\alpha}=  
\sqrt{s_{j}} - M_{\alpha} -\Sigma^{\alpha}_{\eta}(\sqrt{s_{j}}) 
-\Sigma^{\alpha}_{\pi}(\sqrt{s_{j}})-\Sigma^{\alpha}_{\pi\pi}
(\sqrt{s_{j}})\ .  
\label{eq:2.1.12c}
\end{equation}
Here $\alpha$ is a short-hand notation for the quantum numbers 
$(\ell,2T,2j)$ of the isobar resonance $\alpha$.   
The $M_{\alpha}$ is the bare mass of the isobar $\alpha$ and 
$\Sigma^{\alpha}_{\eta},\;
\Sigma^{\alpha}_{\pi}$, and
$\Sigma^{\alpha}_{\pi\pi}$ in eq.(\ref{eq:2.1.12c})
are the self-energies of the isobar $\alpha$ associated, respectively,
with its coupling to the $\eta N, \; \pi N$, and $\pi\pi N$
channels\cite{bhal}. The coupling constants and form factors are denoted
by $g$ and $v$.
At the $\eta N$ threshold, only the $s$-wave $\eta N$ interaction is 
important, which limits the isobar $\alpha$ to    
$N^{*}(1535)$ or $\alpha=(\ell,2T,2j)= (0,1,1)$.  
Clearly, different  models will have different off-shell extensions 
in energy and momenta for $t_{\alpha}(\sqrt{s_{j}},p',p)$. However,
they should all satisfy eq.(\ref{eq:2.1.10c}) in the on-shell limit.   
  
\subsection{Factorization approximation}\label{sec:2.2}
We define the factorization approximation (FA) by taking the $\eta N$
scattering amplitude in eq.(\ref{eq:2.1.6}) out of the ${\bf Q}$
integration at an {\em ad-hoc} fixed momentum $<{\bf Q}>$:
\begin{eqnarray} 
<{\bf k}'\mid V_{FA}\mid {\bf k}> & =  &<{\bf k}', -({\bf k}'+<{\bf Q}>) \mid
t(\sqrt{\overline{s}})_{\eta N\rightarrow\eta N} 
\mid {\bf k},
-({\bf k}+<{\bf Q}>)> \nonumber \\
 & \times & f({\bf k}'-{\bf k})\ ,
\label{eq:2.2.1}
\end{eqnarray}
where
\begin{equation}
f({\bf k}'-{\bf k}) = \sum_{j}\int d{\bf Q}\;\phi_{j}^{*}(-{\bf k}'-{\bf Q})
\phi_{j}(-{\bf k}-{\bf Q})\ ,
\label{eq:2.2.2}
\end{equation}
is the nuclear form factor having the normalization $f(0)=A$.
In eq.(\ref{eq:2.2.1}), $t_{\eta N\rightarrow\eta N}$ is still defined
by the same functional dependences on various momenta and energies as given
by eq.(\ref{eq:2.1.12}), except that $p'$ and $p$ are now determined
from an ad-hoc momentum $<{\bf Q}>$ in the $\eta$-nucleus system 
and that the interaction is given by an ad-hoc 
energy $\sqrt{\overline{s}}$. The choice of $<{\bf Q}>$ is not unique. 
One option is to take an average of two geometries corresponding, respectively,  
to having a motionless target nucleon fixed 
before and after the $\eta N$ interaction.
This leads to
\begin{equation}
<{\bf Q}> = -\;\left ( \frac{A-1}{2A} \right )\;({\bf k}'-{\bf k})\ .
\label{eq:2.2.3}
\end{equation}
This choice has the virtue of preserving the symmetry of the $t$-matrix
with respect to the interchange of ${\bf k}$ and ${\bf k}'$. (There
are other possible schemes; see for example ref.~\cite{cel2}.)
An inspection of eq.(\ref{eq:2.1.7})  suggests that it is  reasonable
to assume $\sqrt{\overline{s}} =M_{\eta}+M_{N}-\Delta \equiv \sqrt{s_{th}} - 
\Delta$, with $\Delta$ being an energy shift parameter. 
In $\pi N$ scattering, the downward shift $\Delta$ that fit the data was 
determined to be $\sim 30$ MeV~\cite{liu1,cott}.

\subsection{``On-shell'' optical potential}\label{sec:2.3}
We will use the term ``on-shell'' optical potential for the optical
potential where on-shell 
hadron-nucleon ($hN$) scattering length is used to generate the $hN$ interaction. 
In the literature, the first-order low-energy 
hadron-nucleus ``on-shell'' optical potential is often 
given as~\cite{sche,stri}
\begin{eqnarray}
 <{\bf k}'\mid U\mid {\bf k}> \;= -\; \frac{1}{4\pi^{2}\mu}
\left ( 1+ \frac{M_{h}}{M_{N}}\right )\; f({\bf k'}-{\bf k}) \nonumber \\
 \times \; \sum_{\ell=0,1} \frac{ {\mid{\bf k}'\mid}^{\ell}
{\mid{\bf k}\mid}^{\ell}}{(1+M_{h}/M_{N})^{2\ell}}
\;a_{hN}^{(\ell)}\ P_{\ell}(\hat{\bf k'}\cdot\hat{\bf k})\ ,
\label{eq:2.3.1}
\end{eqnarray}
where $M_{h}$ is the hadron mass, $\mu$ the hadron-nucleus reduced mass, 
$f$ the nuclear form factor normalized as $f(0)=A$. 
For $\eta$-mesic nuclei, $M_{h}=M_{\eta}$ and 
$a_{\eta N}^{(0)}$
is the $s$-wave $\eta N$ scattering length. 
As has been shown in ref.\cite{hai1}, 
$U$ corresponds to $V_{FA}$ with no energy shift ($\Delta = 0$) and in a static 
limit of the target nucleon. In this latter limit, the $\eta N$ relative 
momenta are ${\bf p}={\bf k}/(1+M_{\eta}/M_{N})$ and  
${\bf p'}={\bf k'}/(1+M_{\eta}/M_{N})$.  
In addition,   $\hat{\bf k}'\cdot{\bf k} = 
\hat{\bf p'}\cdot\hat{\bf p}$.    

\section{Results and discussion} \label{sec:3}
The binding energies and half-widths of $\eta$-mesic nuclei given
by the off-shell microscopic calculation are presented in
table~\ref{table:2}. The covariant optical potential [eq.(\ref{eq:2.1.6})] 
used in the calculation is based on the
model of ref.~\cite{bhal}. We use this model to demonstrate the effect
of subthreshold $\eta N$ interaction on the formation of $\eta$-mesic
nucleus. The solutions are
obtained with the $\eta N$ interaction parameters $g_{\eta N\alpha}$,
$M_{\alpha}$, and $\Lambda_{\eta N\alpha}$ determined
from the $\pi N$ phase shifts of Arndt et al.
The $p$-wave and $d$-wave interactions are also attractive at the threshold  
but their magnitudes are very small and have  
negligible effect on $\epsilon_{\eta}$ and $\Gamma_{\eta}$.
The nuclear wave functions in eq.(\ref{eq:2.1.6}) are derived
from the experimental form factors with the proton finite size
corrected for~\cite{liu4}. Details of the calculation can be
found in refs.~\cite{hai1,hai2}. As can be seen from the table,
the binding energy increases as the nucleus becomes heavier.
In addition, the number of nuclear orbital in which the $\eta$ is bound
($1s,\;2s,\;1p$, etc.) increases with increasing mass number $A$. The  
reason for this trend is discussed in ref.\cite{hai1}.  We would like to
point out that our microscopic calculation does not use on-shell
scattering length as an input. 

We emphasize that the reason we  
use the off-shell model of ref.\cite{bhal} in our comparative analysis     
is because we have at our disposal the detailed unitary 
off-shell momenta and energy dependences of that model, which allow us to 
study the off-shell effects in the formation of
$\eta$-nucleus bound-state. Such off-shell information 
is not readily obtainable from other $\eta N$ models, either because  
the models cannot be extended to off-shell domain or because of the elaborate
computation  required to generate the needed off-shell
$p, p'$, and $\sqrt{s}$ dependences of the $\eta N$ interaction.
Fortunately, as we shall see later, the effects of off-shell dynamics
on any reasonable off-shell model will be qualitatively
similar.  

In table~\ref{table:3}, we present the bound-state solutions obtained from
using the factorized covariant potential $V_{FA}$ [eq.(\ref{eq:2.2.1})] 
with $\Delta = 0,\;10,\;20,\;30$~MeV. The same interaction parameters were 
used. The nuclear form factors~\cite{hofs,jage} used in the
calculations are summarized in table~\ref{table:6}. A comparison
between tables~\ref{table:2} and \ref{table:3} indicates that the FA
results with $\Delta = 30$~MeV are very close to the 
off-shell results. This value of $\Delta$ is similar to the one
found in pion-nucleus elastic scattering~\cite{cott} and can be understood
by noting that the average nuclear binding and Fermi motion amount to
about 30~MeV downward shift~\cite{liu1} of the hadron-nucleon interaction
energy $\sqrt{s}$. Our full off-shell dynamical calculations indicate, 
therefore, that the $\eta N$ interaction in $\eta$ bound state formation  
takes place at energies about 30 MeV below the (free-space) threshold. 

The subthreshold nature of the hadron-nucleon interaction in a nucleus is
also evident in $K$-mesic and $\pi$-mesic atoms. For example, although the 
free-space $KN$ scattering length is repulsive, the effective $KN$ scattering 
length needed to fit the $K$-mesic atom data is attractive\cite{koch}.
The sign change of the scattering length can be easily understood. 
Indeed, one may note that the $\Lambda(1405)$ resonance 
is situated about 26 MeV below the $KN$ 
threshold\cite{koch}. Using a downward shift of 30 MeV found in this work, 
we can see that the in-medium $KN$ interaction occurs actually below the 
resonance. Hence, an attraction arises (see sec.~\ref{sec:1}).    
In the $\pi$-mesic atom studies, 
a strong $s$-wave repulsion of the effective $\pi N$ interaction is 
indicated by the data. This is in sharp contrast to the free-space $\pi N$
$s$-wave interaction where there is a near-perfect cancellation between the 
$S_{31}$ and $S_{11}$ scattering lengths. Bhalerao and Shakin\cite{shak} 
showed that the strong $s$-wave repulsion is due to  
different energy dependences of the $S_{31}$  and $S_{11}$
interactions, which make the cancellation no longer near complete at
subthreshold energies. The mesic-atom experiments, therefore, indicate 
unambiguously that in bound-state formation the 
hadron-nucleon interaction inside a nucleus
occurs at subthreshold energies.  

The results of the ``on-shell'' optical potential  [eq.(\ref{eq:2.3.1})] 
are given in table~\ref{table:4} for two different
values of $a_{\eta N}$. The first one is the same one used for
tables~\ref{table:2} and \ref{table:3}, the other is chosen from
table~\ref{table:1},
such that it has an imaginary part very similar to the former one.
The nuclear form factors are the same as those used in the FA calculations. 
For these two scattering lengths, no bound state can exist in $^{3}$He.
Upon comparing the third column of table~\ref{table:4} with the off-shell  
calculation (table~\ref{table:2}), we can see that 
the ``on-shell'' approximation predicts 
more strongly bound $\eta$-mesic nuclei. 
Also, as expected, the ``on-shell'' 
results for $a_{\eta N}=(0.28+0.19i)$~fm are similar to those of FA with
$\Delta =0$~MeV.
We further note from table~\ref{table:4} 
that the effect on decreasing the calculated 
nuclear binding energies as caused by the 
decrease in ${\cal R}e~a_{\eta N}$ is not a linear function of  
the nuclear mass number $A$. For light nuclei ($A\leq 12$) 
a decrease of ${\cal R}e\;a_{\eta N}$  by a factor of  
$\sim 2$ causes $\epsilon_{\eta}$ to decrease by a
factor greater than 10.

While the in-medium $\eta N$ interaction strength is a function of   
the off-shell variables $p', p, \sqrt{s}$, one can appreciate the main 
feature of the off-shell effects by considering only the $\sqrt{s}$, 
or the energy, dependence. 
Our study of the three different optical
potentials indicates that
this strength decreases with energy.  
However, specific subthreshold energy dependence is model-dependent. 
We may define a phenomenological reduction factor $R$ by  
\begin{equation} 
t_{\eta N\rightarrow \eta N}(\sqrt{s_{th}}-\Delta,p',p)= R(\Delta)
t_{\eta N\rightarrow \eta N}(\sqrt{s_{th}},p',p)\ ,  
\label{eq:4.1.1} 
\end{equation}
and introduce an effective in-medium scattering length by 
\begin{equation} 
a^{eff}_{\eta N} = R(\Delta)\ a_{\eta N}\ . 
\label{eq:4.1.2} 
\end{equation}  
In the model of ref.\cite{bhal}, $R(\Delta) = \sqrt{s_{th}}\  
D(\sqrt{s_{th}})/[(\sqrt{s_{th}}-\Delta) D({\sqrt{s_{th}}-\Delta})]$
[eqs.(\ref{eq:2.1.10}) and (\ref{eq:2.1.12})--(\ref{eq:2.1.12c})].
Our calculation indicates that
$a^{eff}_{\eta N}= (0.26+0.13i),\;(0.25+0.11i),\;(0.23+0.09i) $~fm 
for $\Delta=10,\; 20,\; 30$~MeV, respectively, 
while at the threshold $a_{\eta N}= (0.28+0.19i)$ fm. The $R(\Delta)$ 
is about 0.89 at $\Delta=20$ MeV and
0.82 at $\Delta=30$ MeV. This reduction is the origin of the
$\Delta$-dependence
of the $\epsilon_{\eta}$ and $\Gamma_{\eta}$ in table~\ref{table:3}. For the  
scattering length of the model of ref.\cite{gre3}, it was mentioned that   
at 20 and 30 MeV below the threshold,  
$a^{eff}_{\eta N}= (0.49+0.10i)$~fm and $(0.45+0.08i)$~fm, respectively. 
Upon comparing these values with $a_{\eta N}=(0.75+0.27i)$~fm given by that  
model at the threshold, we see a reduction of more than 1/3 of the 
real part of $a_{\eta N}$ (i.e., $R = 0.6$). It seems that higher is the
${\cal R}e~a_{\eta N}$ at the threshold greater is the subthreshold
reduction. We note that at 20 to 
30 MeV below the threshold, both models give an 
imaginary part of $a^{eff}_{\eta N}$ about $0.09i$ fm.  This is because
the ${\cal I}m ~a^{eff}_{\eta N}$ is related to the reaction channel. At the
subthreshold region, the only reaction channels that the $\eta N$ channel 
can couple to are the $\pi N$ and $\pi\pi N$ channels which were taken into
account by both the models \cite{bhal,gre3}.
On the other hand, the real part of the effective
scattering length is still very model dependent, though in lesser degree 
than at the threshold.  Because of this model dependence, it is not possible
to guess the subthreshold reduction for   
the other models listed in 
table~\ref{table:1} for which the off-shell dependence  
cannot be easily reconstructed from the corresponding 
publications. We conclude from the two models analyzed above that  
a substantial reduction of attractive strength, ${\cal R}e~a_{\eta N}$, 
must occur at subthreshold energies. 

This reduction in attractive strength is of very general nature as it
is a direct consequence
of the $N^{*}(1535)$ resonance. We recall that the
attraction between the $\eta$ and nucleon at low energies arises because
the $\eta N$ threshold is situated below this resonance. However, as the
energy shift $\Delta$ becomes larger, the $\eta N$ interaction energy
moves farther downward away from the resonance. Any reasonable
off-shell model leads, therefore, to a reduced attraction. Consequently,    
calculations making use of free-space scattering length (corresponding to  
$\Delta=~0$) necessarily overestimate $\epsilon_{\eta}$. 
Because of the large sensitivity of the binding energies to $\eta$-nucleon 
interaction in light nuclei and because  
${\cal R}e~a^{eff}_{\eta N}\ll {\cal R}e~a_{\eta N}$, 
many $\eta$-nucleus bound states in very light nuclei, 
as predicted by using some of the $a_{\eta N}$ in table~\ref{table:1}, 
may not exist in real situation. We caution,  
therefore, against using ``on-shell'' (or scattering-length)  
prediction as guide for searching $\eta$-mesic nuclei.

In view of the existing interest in searching for $\eta$-nucleus bound states 
in light nuclear systems, we believe that it is informative to determine   
values of the ``minimal'' scattering length, $a^{min}_{\eta N}$,  
which represents the least value of an $a^{eff}_{\eta N}$ that  
can bind the $\eta$ into an $1s$ nuclear orbital. Clearly, the real and 
imaginary parts of this scattering length are not independent of each other.
We fixed ${\cal I}m~a_{\eta N}^{min}$ to 0.09~fm, as suggested by the two 
off-shell models discussed above, and searched for 
${\cal R}e~a^{min}_{\eta N}$. The results for several light nuclei 
are given in table~\ref{table:5}.

It is interesting to revisit the FSI analysis and to examine the 
results of refs.~\cite{wilk,will} in the light of our findings.  
The FSI analysis was first applied by Wilkin\cite{wilk} to $\eta$-$^{3}$He 
system which is a final state of the $pd\rightarrow \eta^{3}$He reaction. 
The analysis raised a great deal of theoretical interest\cite{wyce} and 
was later extended to the study of $\eta$-$^{4}$He system\cite{will},
which is a final state of the $dd\rightarrow \eta^{4}$He reaction.
Let us denote the $\eta$-nucleus scattering length  
as $A_{0}\equiv A_{R} + i A_{I}$. According to the Watson FSI
theory~\cite{wats}
for a weak transition and a strong FSI, one can approximate 
the total reaction amplitude, $f$, at low energies by   
\begin{equation} 
 \mid f\mid ^{2} = \frac{\mid{f_{B}}\mid^{2}}{\mid 1 - i{A_{0}\peta}\mid^{2}}  
  = \frac{\mid f_{B}\mid^{2}}{(1+A_{I}\peta)^{2} + (A_{R}\peta)^{2}} \ ,  
\label{eq:5.1} 
\end{equation} 
where $\peta$ is the c.m. momentum of the $\eta$ meson. The $f_{B}$ is the
transition amplitude and was treated as a constant in ref.\cite{wilk}. 
The unitarity condition requires $A_{I}>0$. Hence,  
$\mid f\mid^{2}$ increases monotonically when $\peta$ decreases.
A good fit to the $\peta$-dependence of $\mid f\mid^{2}$ was obtained 
by Wilkin who used an $a_{\eta N}=(0.55\pm0.20 + 0.30i)$ fm, which 
led to an $A_{0}(\eta^{3}$He) $= (-2.31 + 2.57i)$~fm~\cite{wilk}.
In a later work, Wilkin and his collaborators\cite{will} 
used $a_{\eta N} \approx (0.52 + 0.25i)$ fm which gave  
$A_{0}(\eta^{3}$He)$\approx (-2.3 + 3.2i)$ fm and
$A_{0}(\eta^{4}$He$)\approx (-2.2 + 1.1i)$ fm,
respectively. Upon introducing these $A_{0}$ into
eq.(\ref{eq:5.1}), they obtained a good representation of the experimental
$\peta$-dependence of $\mid f\mid^{2}$ for both $^{3,4}$He.
Because in the case of a real potential, a negative scattering length 
has the possibility of corresponding either to a repulsive interaction 
or to an attractive 
interaction that supports a $s$-wave bound state \cite{newt},  
the findings of ref.\cite{will} has raised the hope that 
$\eta$-nucleus bound states might exist in $^{3,4}$He.   

A closer look at the value of the $\eta$-helium 
scattering lengths can provide partial answer to the question as to
whether $\eta$ can be bound in $^{3,4}$He nuclei.
This is because for a complex potential the existence of a
bound state imposes a constraint on the  
complex scattering length. Equation~(\ref{eq:5.1}) 
indicates that, if a weakly-bound state exists, 
the amplitude $f$ will have a pole, 
$p_{pol}$, in the complex $\peta$-plane at 
\begin{equation}  
 p_{pol} = \frac{-i}{A_{0}} = \frac{-A_{I}-iA_{R}}{d}\ ,
\label{eq:5.2} 
\end{equation} 
where $d=A_{R}^{2}+A_{I}^{2}$ and $A_{I}$ is positive. 
The condition for a bound state is
${\cal R}e\ p_{pol}^{2} < 0$. This requires that $\mid A_{I}\mid < 
\mid A_{R}\mid$. It is easy to see that the $A_{0}(\eta^{3}$He) 
of ref.\cite{will}  does not satisfy this 
inequality while the $A_{0}(\eta^{4}$He) does. Our  
optical-potential calculations have verified this, namely,
with the $a_{\eta N}$ used in ref.\cite{will} we have found that there is 
no bound state in $^{3}$He but there is one in $^{4}$He. In other words,
$A_{R} < 0$ is not a sufficient condition for having a
bound state. Consequently, in spite of the much greater slope 
of $\mid f\mid^{2}$ with respect to $\peta$ in the $\eta$-$^{3}$He system
\cite{will}, we conclude that there is no $\eta$-nuclear bound state   
in $^{3}$He. For the same reason, $\eta$ cannot be bound onto the
deuteron because none of the various calculated $A_{0} (\eta d)$ of
ref.~\cite{gre1} can satisfy the condition
$\mid A_{I} \mid < \mid A_{R} \mid$.
On the other hand, the formation of a bound state in $^{4}$He
remains a possibility. 

\begin{figure} 
\centerline{\psfig{figure=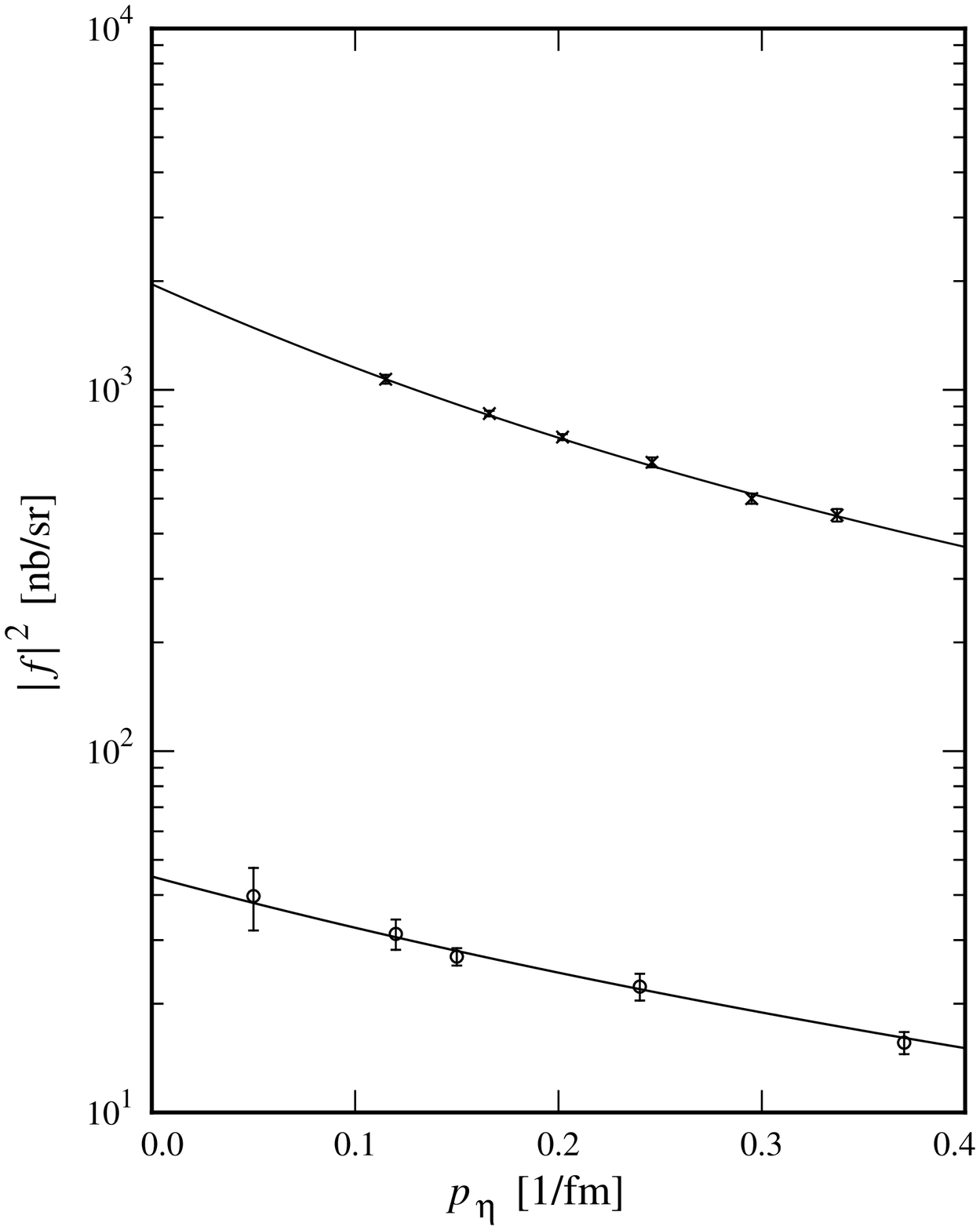,angle=-0,height=5.25in}}   
\caption[dependence]{Dependences of $\mid f\mid^{2}$ on $\peta$. Data for the
$\eta^{3}$He (crosses) and $\eta^{4}$He (open circles) systems  
are from refs.\cite{wyce} and \cite{will}, respectively.  
Solid curves are the results of using the scattering 
lengths having $A_{R} > 0$. }   
\label{fig1}
\end{figure}

This possibility is, however, hampered by the fact that 
$\mid f\mid^{2}$ is insensitive to the sign of $A_{R}$, as can be seen 
from eq.(\ref{eq:5.1}).     
Indeed, we have found an $A_{0}$ that has a positive 
real part and can equally  
describe the data. For example, with $a_{\eta N}= (0.30 + 0.09i)$ fm, 
we have obtained from our optical potential, eq.(\ref{eq:2.3.1}),   
an $A_{0}(\eta^{3}$He)= 2.10 + 2.88i fm which can well describe  
the $^{3}$He data.
Similarly, a good representation of the $^{4}$He data can be obtained with 
an $A_{0}(\eta^{4}$He) $= (0.80 +1.75i)$ fm by using $a_{\eta N} = 
(0.16 +0.09i)$ fm. The quality of representation is shown in fig.\ref{fig1}.
In both cases $A_{R}>0$ which 
corresponds to the situation where the interaction is attractive  
but not strong enough to support a bound state. Hence, we have shown that
FSI can also admit solutions corresponding to having no $\eta$ bound state. 
We emphasize that our above analysis does not imply  
our preference for weaker $a_{\eta N}$ but rather to exemplify numerically 
that FSI cannot provide a unique answer concerning the formation of 
bound state. 

At this point, several comments are in order. Firstly,    
although we need different $a_{\eta N}$ for $^{3}$He and $^{4}$He while in ref.
\cite{will} only one $a_{\eta N}$ was needed, our result is reasonable because
it reflects the  
fact that the transition amplitudes $f_{B}$ in the two reactions 
may not be treated as $\peta$-independent and, thus, do not scale each other by a 
multiplicative constant.   
Indeed, by using the two-step model of ref.\cite{fald},  
Willis {\it et al.}\cite{will} cannot fit the data of   
$\mid f(\peta)\mid^{2}$ for the $\eta$-$^{3}$He and
$\eta$-$^{4}$He systems with the same $a_{\eta N}$. 
In addition, one may note that second-order $\eta$-nucleus optical potential
is important in light nuclei. This potential is sensitive to the nucleon-nucleon 
($NN$) correlation which is very different in $^{3}$He and 
$^{4}$He, as evidenced by the fact that the  
matter densities derived from the measured charge densities of these two nuclei 
have spatial dependences that do not scale each other. Inclusion
of second-order optical potential will lead to different modifications of 
$A_{0}(\eta ^{3,4}$He). Consequently, when only first-order optical potential is
used to generate the $A_{0}$ in fitting the data, the effective $a_{\eta N}$ 
will be different. In a similar manner, the different
$NN$ correlations in the two helium nuclei also    
contribute to the nonscaling of the $f_{B}$. Indeed, many reaction processes
contribute to the transition matrix element. An important subset of the 
processes are those that cannot be approximated    
by the $pp\rightarrow \pi^{+}d$ and $pd\rightarrow$ 
$\pi^{+}\ {^{3}}$He doorway mechanisms~\cite{fald}.
This is because in these processes
the intermediate pion interacts with two different nucleons that do not
belong to the same deuteron cluster in the nucleus. Hence, these processes  
depend strongly on $NN$ correlations having momentum 
dependence that cannot be obtained from using  
the deuteron wave function.     
The non-simple nature of $f_{B}$ is further exemplified by the work  
of Santra and Jain~\cite{sant}, who can describe the
$\eta$-$^{3}$He data
even without FSI when the exchanges of $\rho$ and other mesons 
are included in the calculation of $f_{B}$. While we
strongly believe that FSI must be taken into account, ref.\cite{sant} 
does indicate that there are many more aspects of the dynamics
left to be thoroughly investigated before a definitive conclusion could be   
drawn from the $\eta$-$^{3,4}$He data. 

Secondly, our use of ${\cal I}m\ a^{eff}_{\eta N}$ = 0.09 fm is merely 
suggested by the 30-MeV downward shift 
of the effective $\eta N$ interaction energy, as has been discussed  
after eq.(\ref{eq:4.1.2}). With ${\cal I}m\ a^{eff}_{\eta N}$= 0.09 fm,
the helium data constrain the real part of the effective scattering length to be  
between 0.16 and 0.30 fm. However, we emphasize that  
the value of 0.09 fm is model-dependent and it is important to see if 
different $\eta N$ off-shell models would give a similar 
subthreshold value. Consequently, our finding of repulsive scattering length
for the helium systems should not be interpreted as if  
${\cal I}m\ a^{eff}_{\eta N}$=0.09 fm were the final answer, but rather as 
an example showing the nonuniqueness of the solution to the problem. 
In view of the many needed improvements 
in the modeling of the transition amplitude,
searching a same repulsive effective $a_{\eta N}$ for
both data sets is beyond the scope of this work.

One would certainly like to be able to use the 
$\eta$-$^{3,4}$He data to obtain a more stringent constraint on the 
effective subthreshold $\eta N$ interaction. However, 
we conclude from the above discussion that the currently used FSI analysis
cannot provide an unequivocal answer  
not only because there is no   
simple correspondence between the sign of $A_{R}$ and the existence 
of an $\eta$-nucleus bound state, but also because 
the assumption of a $\peta$-independent $f_{B}$ is uncertain.      
Consequently, our main message is that 
one should not rely on using on-shell scattering length and that    
direct detection of bound states is necessary. 

Finally, we have also examined effects of nuclear form factor on the binding of 
$\eta$. The results for $^{3}$He are presented in table~\ref{table:7}.
An inspection of this last table indicates that the binding of $\eta$ 
is not very sensitive to the finer aspect of a realistic nuclear form factor.
A similar situation has also been noted for $^{4}$He. The 
binding energies and widths of the $\eta$ as given by using  
the 3-parameter Fermi or the Frosch form factor are close to each other. 
The most important effect on the formation of $\eta$-mesic nucleus is, 
therefore, from the subthreshold dynamics of the $\eta$N interaction. 

\section{Conclusions} \label{sec:4}
Our study shows that calculated binding energies and widths of $\eta$-nucleus 
bound states strongly depend on the subthreshold dynamics of the $\eta N$ 
interaction. The results of mesic atoms and the present analysis  
indicate that the average $\eta N$ interaction energy in mesic-nucleus 
formation is below the threshold. What matters for the bound-state formation 
is not the $\eta N$ interaction at the threshold but the effective in-medium 
interaction.  Because the subthreshold behavior of 
$\eta N$ interaction is very model dependent, we believe that it is useful
for theorists to publish not only the $\eta$-nucleon scattering length  
$a_{\eta N}$, but also the corresponding subthreshold values 
as a function of $\Delta$. Before the availability of this information, 
we suggest to look for bound states in nuclear systems much heavier than
those indicated by on-shell scattering length.
 
The FSI analysis represents an interesting approach. Since the analysis 
by itself cannot provide a definitive answer as to whether there is   
an attractive $\eta$-nucleus interaction strong enough to bind the $\eta$,   
more direct measurements such as 
the one used in ref.\cite{sok1} are necessary. 

The downward shift in the effective interaction energy 
can lead to a substantial reduction of the attraction 
of in-medium $\eta$-nucleon interaction with respect to its  
free-space value. Consequently, predictions based upon using free-space
$\eta N$ scattering length inevitably overestimates 
the binding of $\eta$. 
This overestimation of the binding, as revealed by this study,  
has never been taken into account in discussing $\eta$-nuclear bound states. 
One must bear this in mind when using the 
predictions given by such calculations
as guide in searching for $\eta$-nucleus bound states.

\pagebreak
\noindent
\begin{table}
\caption{Eta-nucleon $s$-wave scattering lengths $a_{\eta N}$.}
\label{table:1}

\bigskip
\noindent
\begin{tabular}{cll}  
$a_{\eta N}$ (fm) & Formalism/Reaction & Reference \\ 
\tableline
$0.270 + 0.220i$ & Isobar model & Bhalerao and Liu~\cite{bhal} \\
$0.280 + 0.190i$ & Isobar model & {\it ibid} \\
$0.281 + 0.360i$ & Photoproduction of $\eta$ & Krusche~\cite{krus}       \\
$0.430 + 0.394i$ &  & {\it ibid} \\
$0.579 + 0.399i$ &  & {\it ibid} \\
$0.476 + 0.279i$ & Electroproduction of $\eta$ & Tiator et al.~\cite{tiat} \\
$0.500 + 0.330i$ & $pd\rightarrow ^{3}$He$\;e\eta$ & Wilkin~\cite{wilk} \\
$0.510 + 0.210i$ & Isobar model & Sauermann et al.~\cite{saue} \\
$0.550 + 0.300i$ &  & {\it ibid} \\
$0.620 + 0.300i$ & Coupled $T$-matrices & Abaev and Nefkens~\cite{abae} \\
$0.680 + 0.240i$ & Effective Lagrangian &  Kaiser et al.~\cite{kais}\\
$0.750 + 0.270i$ & Coupled $K$-matrices & Green and Wycech~\cite{gre3} \\
$0.870 + 0.270i$ & Coupled $K$-matrices & Green and Wycech~\cite{gre2} \\
$1.050 + 0.270i$ &  & {\it ibid} \\
$0.404 + 0.343i$ & Coupled $T$-matrices & Batini\'{c} et al.~\cite{bat1} \\
$0.876 + 0.274i$ &   & Batini\'{c} and \v{S}varc ~ \cite{bat3} \\
$0.886 + 0.274i$ &   & {\it ibid} \\
$0.968 + 0.281i$ &   & Batini\'{c} et al. ~ \cite{bat2} \\
$0.980 + 0.370i$ & Coupled $T$-matrices & Arima et al.~\cite{arim} \\
\end{tabular}
\end{table}

\begin{table}
\caption{Binding energies and half-widths (both in MeV) of $\eta$-mesic
nuclei given by the full off-shell calculation. The solutions 
were obtained with the $\eta N$ interaction parameters
determined from the $\pi N$ phase shifts of Arndt et al. No bound state
solutions of eq.(1) were found for $A<12$. }

\label{table:2}

\bigskip
\noindent
\begin{tabular}{ccc}  
Nucleus & Orbital ($n\ell$) & $\epsilon_{\eta}+ i\Gamma_{\eta}/2$ \\ 
\tableline
$^{12}$C  &  $1s$ & $ -(1.19 + 3.67i)$   \\
$^{16}$O  &  $1s$ & $ -(3.45 + 5.38i)$   \\
$^{26}$Mg &  $1s$ & $ -(6.39 + 6.60i)$   \\
$^{40}$Ca &  $1s$ & $ -(8.91 + 6.80i)$   \\
$^{90}$Zr &  $1s$ & $-(14.80 + 8.87i)$   \\
          &  $1p$ & $ -(4.75 + 6.70i)$   \\
$^{208}$Pb&  $1s$ & $-(18.46 + 10.11i)$  \\
          &  $2s$ & $ -(2.37 + 5.82i)$   \\
          &  $1p$ & $-(12.28 + 9.28i)$   \\
          &  $1d$ & $ -(3.99 + 6.90i)$   \\
\end{tabular}
\end{table}
\noindent
\begin{table}
\caption{Binding energies and half-widths (both in MeV) of $\eta$-mesic
nuclei obtained with the factorization approach for different
values of the energy shift parameter $\Delta$ (in MeV).}
\label{table:3}

\bigskip
\noindent
\begin{tabular}{cccccc}  
Nucleus & Orbital ($n\ell$) & $\Delta = 0$ & $\Delta = 10$ & $\Delta = 20$
& $\Delta = 30$ \\
\tableline
$^{12}$C  &  $1s$ & $ -(2.18 + 9.96i) $  &  $ -(1.80 + 6.80i)$
& $-(1.42 + 5.19i)$ & $-(1.10 + 4.10i)$ \\
$^{16}$O  &  $1s$ & $ -(4.61 + 11.57i)$  &  $ -(3.92 + 8.13i)$
& $-(3.33 + 6.37i)$ & $-(2.84 + 5.17i)$ \\
$^{26}$Mg &  $1s$ & $ -(10.21 + 15.41i)$  &  $ -(8.95 + 11.17i)$
& $-(7.94 + 8.97i)$ & $-(7.11 + 7.46i)$ \\
$^{40}$Ca &  $1s$ & $ -(14.34 + 17.06i)$  &  $ -(12.75 + 12.55i)$
& $-(11.53 + 10.21i)$ & $-(10.51 + 8.59i)$ \\
$^{90}$Zr &  $1s$ & $-(21.32 + 18.59i)$  &  $-(19.15 + 13.97i)$
& $-(17.58 + 11.54i)$ & $-(16.29 + 9.84i)$ \\
&  $1p$ & $ -(8.27 + 16.01i)$  &  $ -(7.19 + 11.47i)$
& $-(6.23 + 9.48i)$ & $-(5.40 + 7.94i)$ \\
$^{208}$Pb&  $1s$ & $-(24.06 + 19.18i)$ &  $-(21.88 + 14.44i)$
          & $-(20.28 + 11.96)$ & $-(18.96 + 10.22i)$ \\
          &  $2s$ & $ -(4.89 + 11.04i)$  &  $ -(3.67 + 8.28i)$
          & $-(2.81 + 6.79i)$ & $-(2.12 + 5.72i)$ \\
          &  $1p$ & $-(18.33 + 18.97i)$  &  $ -(16.31 + 14.27i)$
          & $-(14.81 + 11.79i)$ & $-(13.56 + 10.06i)$ \\
          &  $1d$ & $ -(8.27 + 14.07i)$  &  $ -(6.17 + 10.56i)$
          & $-(5.58 + 8.71i)$ & $-(4.66 + 7.41i)$ \\
\end{tabular}
\end{table}

\begin{table}
\caption{Nuclear form factors used in the factorization approach 
and scattering-length approach.}
\label{table:6}

\bigskip
\noindent
\begin{tabular}{cll}
Nucleus & Form factor & Parameters\tablenotemark[1] \\
\tableline
$^{3}$He  &  Hollow Exponential & $a=1.82$~fm \\
          & Gaussian &      $a=1.77$~fm \\
$^{4}$He  &  3--parameter Fermi & $c=1.01$~fm, $\;z=0.327$~fm,
                                  $\;w=0.445$~fm\\
          & Frosch model   & $a=0.316$~fm, $b=0.680$~fm \\                             
$^{6}$Li  &  Modified Harmonic Well & $a_{1}=1.71$~fm, $a_{2}=2.08$~fm \\
$^{9}$Be  &  Harmonic Well & $\alpha =2/3,\;\;a=2.42$~fm \\
$^{10}$B  &  Harmonic Well & $\alpha = 1,\;\;a=2.45$~fm \\
$^{11}$B  &  Harmonic Well & $\alpha = 1,\;\;a=2.42$~fm \\
$^{12}$C  &  Harmonic Well & $\alpha = 4/3,\;\;a=2.53$~fm \\
$^{16}$O  &  Harmonic Well & $\alpha = 1.6,\;\;a=2.75$~fm \\
$^{26}$Mg &  2--parameter Fermi & $c=3.050$~fm, $\;z=0.524$~fm \\
$^{40}$Ca &  2--parameter Fermi & $c=3.510$~fm, $\;z=0.563$~fm \\
$^{90}$Zr &  3--parameter Gaussian & $c=4.500$~fm, $\;z=2.530$~fm,
                                     $\;w=0.20$~fm\\
$^{208}$Pb&  2--parameter Fermi & $c=6.624$~fm, $\;z=0.549$~fm \\
\end{tabular}
\tablenotemark[1]
{Ref.~\cite{hofs} for $A=3-16$ and ref.~\cite{jage} for the rest of the
nuclei.}
\end{table}

\noindent
\begin{table}
\caption{Binding energies and half-widths (both in MeV) of $\eta$-mesic
nuclei given by the scattering-length approach    
for two different values of the scattering length $a_{\eta N}$. A blank
entry indicates the absence of bound state. No bound state exists in 
$^{3}$He.}
\label{table:4}

\bigskip
\noindent
\begin{tabular}{cccc}  
Nucleus & Orbital ($n\ell$) & $a_{\eta N}=(0.28+0.19i)$~fm &
$a_{\eta N}=(0.51+0.21i)$~fm \\
\tableline
$^{4}$He\tablenotemark[2]& $1s$ &     & $-(6.30+11.47i)$   \\
$^{6}$Li  &  $1s$ &                   & $-(3.47+6.79i)$    \\
$^{9}$Be  &  $1s$ &                   & $-(13.78+12.45i)$  \\
$^{10}$B  &  $1s$ & $-(0.93+8.70)$    & $-(15.85+13.05i)$  \\
$^{11}$B  &  $1s$ & $-(2.71+10.91i)$  & $-(20.78+15.42i)$  \\
$^{12}$C  &  $1s$ & $-(2.91+10.22i)$  & $-(19.61+14.20i)$  \\
$^{16}$O  &  $1s$ & $-(5.42+11.43i)$  & $-(23.26+14.86i)$  \\
          &  $1p$ &                   & $-(0.95+7.72i)$    \\
$^{26}$Mg &  $1s$ & $-(11.24+14.76i)$ & $-(33.11+17.73i)$  \\
          &  $1p$ &                   & $-(13.41+12.33i)$  \\
$^{40}$Ca &  $1s$ & $-(15.46+16.66i)$ & $-(38.85+19.16i)$  \\
          &  $2s$ &                   & $-(5.59+6.14i)$    \\
          &  $1p$ & $-(1.22+10.58i)$  & $-(22.84+14.32i)$  \\
          &  $1d$ &                   & $-(4.28+9.52i)$    \\
$^{90}$Zr &  $1s$ & $-(22.41+19.97i)$ & $-(48.40+22.60i)$  \\
          &  $2s$ &                   & $-(26.07+10.07i)$  \\
          &  $1p$ & $-(10.18+14.33i)$ & $-(31.53+15.93i)$  \\
          &  $2p$ &                   & $-(18.51+8.57i)$   \\
$^{208}$Pb&  $1s$ & $-(24.55+19.57i)$ & $-(50.27+21.42i)$  \\
          &  $2s$ & $-(10.56+13.32i)$ & $-(22.27+11.50i)$  \\
          &  $1p$ & $-(20.19+19.05i)$ & $-(34.03+10.03i)$  \\
          &  $2p$ &                   & $-(1.89+3.75i)$    \\
          &  $1d$ & $-(12.22+16.07i)$ & $-(27.89+12.17i)$  \\
\end{tabular}
\tablenotemark[2]
{Form factor used is the 3-parameter Fermi.}
\end{table}

\noindent
\begin{table}
\caption{Values of $a^{min}_{\eta N}$ for 
nuclei with mass number $A<10$. The ${\cal I}m~a^{min}_{\eta N}$ 
is fixed at 0.09 fm (see the text).}  
\label{table:5}

\bigskip
\noindent
\begin{tabular}{clc}  
Nucleus & Nuclear form factor & $a^{min}_{\eta N}$~(fm) \\  
\tableline
$^{3}$He  & Hollow exponential & $0.49+0.09i$   \\
$^{4}$He  & 3-parameter Fermi  & $0.35+0.09i$   \\
$^{6}$Li  & Modified harmonic well & $0.35+0.09i$ \\ 
$^{9}$Be  & Harmonic Well &  $0.24+0.09i$   \\
\end{tabular}
\end{table}

\bigskip
\noindent
\begin{table}
\caption{Dependence of binding energies and half-widths (both in MeV)
of $\eta$-$^{3}$He bound state on two different form factors
(hollow exponential and Gaussian) for a few values of the scattering
length $a_{\eta N}$. All bound states are in the $1s$ orbital.}
\label{table:7}

\bigskip
\noindent
\begin{tabular}{ccc}
$a_{\eta N}$~fm & Hollow Exponential & Gaussian \\
\tableline
$0.680+0.240i$ &  $-(3.74+7.89i) $  &  $-(4.16+7.93i)$ \\
$0.750+0.270i$ &  $-(6.24+9.94i) $  &  $-(6.79+9.94i)$ \\
$0.876+0.274i$ &  $-(11.86+11.26i)$ &  $-(12.59+11.10i)$ \\
\end{tabular}
\end{table}

\newpage 


\bigskip
\begin{references}
 \bibitem{hai1} Q. Haider and L.C. Liu, Phys. Lett. {\bf B 172}, 257
 (1986); {\bf B 174}, 465E (1986).
 \bibitem{bhal}   R.S. Bhalerao and L.C. Liu, Phys. Rev. Lett.
 {\bf 54}, 865 (1985).
 \bibitem{liku} G.L. Li, W.K. Cheng, and T.T.S. Kuo,
 Phys. Lett. {\bf B 195}, 515 (1987).
 \bibitem{chri} R.E. Chrien et al., Phys. Rev. Lett. {\bf 60},
 2595 (1988).
 \bibitem{lieb} B.J. Lieb and L.C. Liu, {\it Progress at LAMPF},
 LA-11670-PR (1988).
 \bibitem{john} J.D. Johnson et al., Phys. Rev. {\bf  C 47},
 2571 (1993).
 \bibitem{hai2} L.C. Liu and Q. Haider, Phys. Rev. {\bf C 34},
 1845 (1986).
 \bibitem{hai3} Q. Haider and L.C. Liu, Phys. Rev. {\bf C 36},
 1636 (1987).
 \bibitem{sok1} G.A. Sokol et al., Particles and Nuclei Letters
 {\bf 5}, 71 (2000); see also {\it Proceedings of the IX International
 Seminar on Electromagnetic
 Interaction of Nuclei at Low and Medium Energies,} Moscow, 2000, p.214.
 \bibitem{sok2} G.A. Sokol and L.N. Pavlyuchenko, Los Alamos Archive
 nucl-ex/0111020, November 2001.
 \bibitem{wats} M.L. Goldberger and K.M. Watson, {\it Collision Theory}
 (Wiley, N.Y., 1964).
 \bibitem{gre3} A.M. Green and S. Wycech, Phys. Rev. {\bf C 55},
 R2167 (1997).
 \bibitem{gre2} A.M. Green and S. Wycech, Phys. Rev. {\bf C 60},
 035208 (1999).
 \bibitem{saue} Ch. Sauerman, B.L. Friman, and W. N\"{o}renberg,
 Phys. Lett.  {\bf B 341}, 261 (1995); {\bf B 409}, 51 (1997).
 \bibitem{benn} C. Bennhold and H. Tanabe, Nucl. Phys. {\bf A 530},
 625 (1991).
 \bibitem{abae} V.V. Abaev and B.M.K. Nefkens, Phys. Rev. {\bf C 53},
 385 (1996).
 \bibitem{kais} N. Kaiser, P.B. Siegel, and W. Weise, Phys. Lett.
 {\bf B 362}, 23 (1995).
 \bibitem{bat1} M. Batini\'{c}, I. \v{S}laus, A. \v{S}varc,
 and B.M.K. Nefkens, Phys. Rev. {\bf C 51}, 2310 (1995).
 \bibitem{bat3} M. Batini\'{c} and A. \v{S}varc, Few-Body Systems, {\bf 20}, 
 69 (1996). 
 \bibitem{bat2} M. Batini\'{c}, I. \v{S}laus, and A. \v{S}varc,
 Phys. Rev. {\bf C 52}, 2188 (1995).
 \bibitem{arim} M. Arima, K. Shimizu, and  K. Yazaki, Nucl. Phys.
 {\bf A 543}, 613 (1992).
 \bibitem{tiat} L. Tiator, C. Bennhold, and S. Kamalov, Nucl. Phys.
 {\bf A 580}, 455 (1994).
 \bibitem{krus} B. Krusche, {\it Proceedings of the II TAPS Workshop,}
 Guadamar, 1993, edited by J. Diaz and Y. Schutz (World Scientific,
 Singapore, 1994).
 \bibitem{wilk} C. Wilkin, Phys. Rev. {\bf C 47}, R938 (1993).
 \bibitem{wyce} S. Wycech, A.M. Green, and J.A. Niskanen,
 Phys. Rev. {\bf C 52}, 544 (1995).
 \bibitem{gre1}A.M. Green, J.A. Niskanen, and S. Wycech, Phys. Rev. 
  {\bf C 54}, 1970 (1996).
 \bibitem{rak1} S.A. Rakityanski, S.A. Sofianos, W. Sandhas, and
 V.B. Belayev, Phys. Lett. {\bf B 359}, 33 (1995).
 \bibitem{rak2} S.A. Rakityansky, S.A. Sofianos, M. Braun, V. B. Belayev,
 and W. Sandhas, Phys. Rev. {\bf C 53}, R2043 (1996).
 \bibitem{liu1} L.C. Liu and C.M. Shakin, Prog. Part. and Nucl. Phys.
 {\bf 5}, 207 (1980).
 \bibitem{koch} J.H. Koch and M.M. Sternheim, Phys. Rev. Lett. {\bf 28},
 1061 (1972).
 \bibitem{seki} R. Seki and K. Wiegand, Ann. Rev. Nucl. Sci. {\bf 28},
 241 (1975).
 \bibitem{kwon} Y.R. Kwon and F.B. Tabakin, Phys. Rev. {\bf C 18},
 932 (1978).
 \bibitem{cel1} L.S. Celenza, M.K. Liou, L.C. Liu, and C.M. Shakin,
 Phys. Rev. {\bf C 10}, 398 (1974). Use the $g_{2}$-reduction with 
 $L = 0$. Note that there is a misprint in the first line of eq.(12),
 where $W'$ should have been $2W'-W$. 
 \bibitem{cel2} L. Celenza, L.C. Liu, and C.M. Shakin, Phys. Rev
 {\bf C 12}, 1983 (1975).
 \bibitem{cott} W.B. Cottingame and D.B. Holtkamp, Phys. Rev. Lett.
 {\bf 45}, 1828 (1980). 
 \bibitem{sche} T.E.O. Ericson and F. Scheck, Nucl. Phys. {\bf B 19}, 
 450 (1970); See eq.(18) given therein. 
 \bibitem{stri} K. Stricker, H. McManus, and J.A. Carr, Phys. Rev. {\bf C 19},
 929 (1979); See the kinematically correct potential, eq.(7), given therein.
 \bibitem{liu4} L.C. Liu and C.M. Shakin, Nuovo Cimento {\bf A 53},
 142 (1979).
 \bibitem{hofs} H.R. Collard, L.R.B. Elton, and R. Hofstader,
 {\it Numerical Data and Functional Relationships in Science and
 Technology,} Vol.~2 Nuclear Radii, edited by H. Schopper
 (Springer-Verlag, N.Y., 1967).
 \bibitem{jage} C.W. de Jager, H. de Vries, and C. de Vries,
 Atomic Data and Nuclear Data Tables {\bf 14}, 479 (1974).
 \bibitem{shak} R.S. Bhalerao and C.M. Shakin, Phys. Rev. {\bf C 23}, 
 2198 (1981).
 \bibitem{will} N. Willis et al., Phys. Lett. {\bf B 406}, 14 (1997).  
 \bibitem{newt} Roger G. Newton, {\it Scattering Theory of Waves and 
 Particles} ( Springer-Verlag, N. Y, 1982). The scattering length 
 defined in this reference has an opposite sign to the one used in 
 this work. This sign difference has been taken into account in our 
 discussion.    
 \bibitem{fald} G. F\"{a}ldt and C. Wilkin, Nuc. Phys. {\bf A 596},
 488 (1996); {\bf A 587}, 769 (1995).
 \bibitem{sant} A.B. Santra and B.K. Jain, Phys. Rev. {\bf C 64}, 025201
 (2001).
    
\end{references}
\end{document}